\documentclass[preprintnumbers,article,amsmath,amssymb,floatfix,10pt,prd,twocolumn,
superscriptaddress,nofootinbib]{revtex4-2}
\usepackage{bm}
\usepackage{amsfonts}
\usepackage{latexsym}
\usepackage[latin1]{inputenc}
\usepackage{graphicx}
\usepackage{amsmath}
\usepackage{palatino}
\usepackage{mathpazo}
\usepackage{textcomp}
\usepackage{float}
\usepackage{booktabs}
\usepackage{dcolumn}
\usepackage{ragged2e}
\usepackage{hyperref}
\hypersetup{colorlinks,citecolor=blue}
\hypersetup{colorlinks=true,linkcolor=red,filecolor=magenta,    urlcolor=blue}
\usepackage{amsmath}
\usepackage{xcolor}
\usepackage{orcidlink}
\usepackage{epsfig}
\usepackage{caption}
\usepackage{subcaption}
\usepackage{commath}
\def\jnl@style{\it}
\def\aaref@jnl#1{{\jnl@style#1}}

\def\aaref@jnl#1{{\jnl@style#1}}

\def\aj{\aaref@jnl{AJ}}                   % Astronomical Journal
\def\apj{\aaref@jnl{ApJ}}                 % Astrophysical Journal
\def\apjl{\aaref@jnl{ApJ}}                % Astrophysical Journal, Letters
\def\apjs{\aaref@jnl{ApJS}}               % Astrophysical Journal, Supplement
\def\apss{\aaref@jnl{Ap\&SS}}             % Astrophysics and Space Science
\def\aap{\aaref@jnl{A\&A}}                % Astronomy and Astrophysics
\def\aapr{\aaref@jnl{A\&A~Rev.}}          % Astronomy and Astrophysics Reviews
\def\aaps{\aaref@jnl{A\&AS}}              % Astronomy and Astrophysics, Supplement
\def\mnras{\aaref@jnl{Mon.~Not.~Roy.~Astron.~Soc.}}             % Monthly Notices of the RAS
\def\prd{\aaref@jnl{Phys.~Rev.~D}}        % Physical Review D
\def\prc{\aaref@jnl{Phys.~Rev.~C}}  % Physical Review C
\def\prl{\aaref@jnl{Phys.~Rev.~Lett.}}    % Physical Review Letters
\def\qjras{\aaref@jnl{QJRAS}}             % Quarterly Journal of the RAS
\def\skytel{\aaref@jnl{S\&T}}             % Sky and Telescope
\def\ssr{\aaref@jnl{Space~Sci.~Rev.}}     % Space Science Reviews
\def\zap{\aaref@jnl{ZAp}}                 % Zeitschrift fuer Astrophysik
\def\nat{\aaref@jnl{Nature}}              % Nature
\def\aplett{\aaref@jnl{Astrophys.~Lett.}} % Astrophysics Letters
\def\apspr{\aaref@jnl{Astrophys.~Space~Phys.~Res.}} % Astrophysics Space Physics Research
\def\physrep{\aaref@jnl{Phys.~Rep.}}      % Physics Reports
\def\physscr{\aaref@jnl{Phys.~Scr}}       % Physica Scripta
\def\commat{\aaref@jnl{Comm.~Math.~Phys.}}              % Communications in Mathematical Physics
\def\science{\aaref@jnl{Science}}               % Science
\def\cqg{\aaref@jnl{Classical Quant.~Grav.}}            % Classical and Quantum Gravity
\def\jpcs{\aaref@jnl{JPCS}}                                     % Journal of Physics Conference Series
\def\ijmpd{\aaref@jnl{Int.~J.~Mod.~Phys.~D}}                    % International Journal of Modern Physics D
\def\grg{\aaref@jnl{Gen.~Relat.~Gravit.}}               % General Relativity and Gravitation
\def\rpp{\aaref@jnl{Rep.~Prog.~Phys.}}          % Reports on Progress in Physics
\def\npa{\aaref@jnl{Nucl.~Phys.~A}}        % Nuclear Physics A
\def\lrr{\aaref@jnl{Living Rev.~Rel.}}                   % Living reviews in relativity
\def\jcap{\aaref@jnl{J.~Cosmology Astropart.~Phys.}}    % Journal of cosmology and astroparticle physics
\def\rmp{\aaref@jnl{Rev.~Mod.~Phys.}}   %Reviews of modern physics
\def\epjc{\aaref@jnl{Eur.~Phys.~J.~C}} 
\def\plb{\aaref@jnl{~Phy.~Lett.~B}} 
\def\mpla{\aaref@jnl{Mod.~Phy.~Lett.~A}} 
\def\arxiv{\aaref@jnl{arxiv.org}}

\allowdisplaybreaks[1]
\begin{document}
%\color{red}
\color{black}       %% For one column
\title{Squared torsion $f(T,\mathcal{T})$ gravity and its cosmological implications}

\author{Simran Arora \orcidlink{0000-0003-0326-8945}}
\email{dawrasimran27@gmail.com}
\affiliation{Department of Mathematics, Birla Institute of Technology and
Science-Pilani,\\ Hyderabad Campus, Hyderabad-500078, India.}
\author{Aaqid Bhat}
\email{aaqid555@gmail.com}
\affiliation{Department of Mathematics, Birla Institute of Technology and
Science-Pilani,\\ Hyderabad Campus, Hyderabad-500078, India.}
\author{P.K. Sahoo\orcidlink{0000-0003-2130-8832}}
\email{pksahoo@hyderabad.bits-pilani.ac.in}
\affiliation{Department of Mathematics, Birla Institute of Technology and
Science-Pilani,\\ Hyderabad Campus, Hyderabad-500078, India.}

%%%%%%%%%%%%%%%%%%%%%%%%%%%%%%%%%%%%%  DATE  %%%%%%%%%%%%%%%%%%%%%%%%%%%%%%%%%%%%
\date{\today}

\begin{abstract}

We present the coupling of the torsion scalar $T$ and the trace of energy-momentum tensor $\mathcal{T}$, which produces new modified $f(T,\mathcal{T})$ gravity. Moreover, we consider the functional form $f(T,\mathcal{T}) =\alpha \mathcal{T}+\beta T^2$ where $\alpha$  and $\beta$  are free parameters. As an alternative to a cosmological constant, the $f(T,\mathcal{T})$ theory may offer a theoretical explanation of the late-time acceleration. The recent observational data to the considered model especially the bounds on model parameters is applied in detail. Furthermore, we analyze the cosmological behavior of the deceleration, effective equation of state and total equation of state parameters. However, it is seen that the deceleration parameter depicts the transition from deceleration to acceleration and the effective dark sector shows a quintessence-like evolution.\\


\end{abstract}

\maketitle

%\date{\today}

%%%%%%%%%%%%%%%%%%%%%%%%%%%%%%%%%%%%%%%%%%%%%%%%%%%%%%%%%%%%%%%%%%%%%%%%
%%%%%%%%%%%%%%%        Introduction        %%%%%%%%%%%%%%%%%%%%%%%%%%%%%
%%%%%%%%%%%%%%%%%%%%%%%%%%%%%%%%%%%%%%%%%%%%%%%%%%%%%%%%%%%%%%%%%%%%%%%%
\section{Introduction}
The in-depth verification of late time acceleration has led to immense research towards its explanation. It is commonly known by the observations of type Ia Supernovae \cite{Riess/1998,Perl/1999}, BAO \cite{Eisenstein/2005, Percival/2007}, CMB \cite{Komatsu/2011}, and $H(z)$ measurements \cite{Farooq/2017}.  The dark energy, which tried to explain the late-time acceleration as the outcome of a type of energy connected to the cosmological constant, is one of the successful primary models. In order to navigate the path beyond the typical dark energy models, one can go beyond the general theory of relativity by modifying the geometry. Alternative theories such as $f(R)$ gravity \cite{Staro/2007,Capo/2008,Chiba/2007}, a coupling between matter and curvature through $f(R,\mathcal{T})$ gravity \cite{Harko/2011,Moraes/2017}, where $\mathcal{T}$ is the trace of energy momentum tensor, $f(R,G)$ \cite{Laurentis/2015,Gomez/2012} ($G$ is the Gauss-Bonnet) have all attempted to explain the dark energy phenomenon in the context of curvature.\\
As a result, more general geometries than the Riemannian, which may be valid at solar system level,  may provide an explanation for the behavior of matter at large scales in the universe. There has been a  rising interest in teleparallel gravity, a different type of modified gravity that uses torsion instead of curvature. The basic idea behind the teleparallel approach is to replace the metric of spacetime by a set of tetrad vectors which is the physical variable describing the gravitational properties. Moreover, this mathemcatical development employs a different connection known as the Weitzen$\ddot{o}$ck connection. \\
When one extends the action of the modified gravity based on torsion, a separate and intriguing class of modified gravity arises named as the teleparallel equivalent of general relativity or $f(T)$ gravity. However, a number of analyses in $f(T)$ gravity such as cosmological solutions \cite{Paliathanasis/2016}, late time acceleration \cite{Myrzakulov/2011, Bamba/2011}, thermodynamics \cite{Salako/2013}, cosmological perturbations \cite{Chen/2011}, cosmography \cite{Capozziello/2011} have been applied in the literature.  For a thorough analysis of $f(T)$ gravity, one can check \cite{Cai/2016}.\\
Another new suggestion in modified gravity is to employ the coupling between the torsion and trace of energy-momentum tensor known as $f(T,\mathcal{T})$ theory, in a similar fashion as $f(R,T)$ gravity. The $f(T,\mathcal{T})$ gravity has been proposed in \cite{Harko/2014}, and its consistency with cosmological data and the necessary physical conditions for a coherent cosmological theory still has to be validated. The coupling of torsion and matter expands the possibilities for describing the characteristics of dark energy or, more specifically, what is driving the observed acceleration. This theory has been investigated in the context of reconstruction and stability \cite{Junior/2016,Momeni/2014}, late-time acceleration and inflationary phases \cite{Harko/2014}, growth factor of sub-horizon modes \cite{Farrugia/2016}, quark stars \cite{Pace/2017}. \\
The goal of the current study is to construct the extended coupled-matter modified gravity by starting with TEGR rather than GR. The construction of $f(T,\mathcal{T})$ gravity, which allows for arbitrary functions of the torsion scalar $T$ and the trace of the energy-momentum tensor $\mathcal{T}$, is the focus of the current effort. In this paper, we investigate a squared-torsion $f(T,\mathcal{T})$ model, that raises a question on the viability of such a theory as a candidate to account for late-time acceleration. Further, the parameters are constrained using the set of observational datasets and in particular, we check the late-time accelerating behavior holds true for $f(T,\mathcal{T})$ using the cosmological parameters. \\
The plan of the work is the following: Starting from the background of $f(T)$, we introduce the framework of $f(T,\mathcal{T})$ gravity in section \ref{sec2}. Section \ref{sec3} is devoted to the cosmological framework and the solutions to the field equations. Specifically, in section \ref{sec4}, we deal with the observational data and methodology used to constrain the parameters involved. 
The late-time accelerated phase is examined in section \ref{sec5} through cosmological evolution. Finally, a conclusion is given in section \ref{sec6}.

\section{Field equations} \label{sec2}
The fundamental preliminaries for the reconstruction of the $f(T)$ and $f(T,\mathcal{T})$ theories of gravity are presented in this section.\\
One needs a new connection, the $Weitzenb\ddot{o}ck$ connection \cite{Aldrovandi/2013} to obtain the torsion-based theory defined as $\tilde{\Gamma}^{\alpha}_{\mu \nu}= e_{a}^{\,\, \alpha} \partial_{\nu} e^{a}_{\,\,\mu}$, where $e^{a}_{\,\, \mu}$ and $e_{a}^{\,\, \alpha}$ are tetrads (or vierbeins). These vierbeins relate to the metric tensor $g_{\mu \nu}$ at each point x of the spacetime manifold as
\begin{equation}
\label{1}
g_{\mu \nu}(x)= e^{a}_{\,\, \mu}(x) e^{b}_{\,\, \nu}(x) \eta_{ab}.
\end{equation}
Here, $\eta_{ab}= diag(1,-1,-1,-1)$ is the Minkowski metric tensor.
Hence, the torsion tensor describing the gravitational field is 
\begin{equation}
\label{2}
T^{\alpha}_{\,\,\mu \nu}= \Gamma^{\alpha}_{\,\,\nu \mu}- \Gamma^{\,\,\alpha}_{\mu \nu}= e_{a}^{\,\,\alpha}\left(\partial_{\mu} e^{a}_{\,\,\nu}-\partial_{\nu} e^{a}_{\,\,\mu} \right).
\end{equation}
We define the contortion and the superpotential tensor through the components of the torsion tensor
\begin{eqnarray}
\label{3}
K^{\mu \nu}_{\,\, \alpha} &=& -\frac{1}{2}\left(T^{\mu \nu}_{\,\, \alpha} - T^{\nu \mu}_{\alpha} - T_{\alpha}^{\,\, \mu \nu}   \right),\\
\label{4}
S_{\alpha}^{\,\, \mu \nu} &=& \frac{1}{2}\left(K^{\mu \nu}_{\,\, \alpha} + \delta^{\mu}_{\alpha} T^{\lambda \mu}_{\,\, \lambda}- \delta^{\nu}_{\alpha} T^{\lambda \mu}_{\,\, \lambda}   \right).
\end{eqnarray}
Using equations \eqref{2} and \eqref{4}, one obtain the torsion scalar \cite{Harko/2014,Maluf/2013,Cai/2016}
\begin{equation}
\label{5}
T = S_{\alpha}^{\,\, \mu \nu} T^{\alpha}_{\,\,\mu \nu}= \frac{1}{2} T^{\alpha \mu\nu} T_{\alpha \mu\nu} + \frac{1}{2} T^{\alpha \mu\nu} T_{ \nu\mu \alpha} - T_{\alpha \mu}^{\,\, \,\, \alpha} T^{\nu \mu}_{\,\,\,\, \nu} .
\end{equation}

One can define the gravitational action for teleparallel gravity by
\begin{equation}
\label{6}
S= \int d^4 x\, e\, [ T+ \mathcal{L}_{m}],
\end{equation}
where $e=det(e^{a}_{\,\,\mu})=\sqrt{-g}$ and $\mathcal{L}_{m}$ is the matter Lagrangian. In fact, one can extend $T$ to $T+f(T)$, the so called $f(T)$ gravity. Moreover, it can be generalized to become a general function of both the torsion scalar and the trace of the energy-momentum tensor $\mathcal{T}$, which results in the $f(T,\mathcal{T})$ gravity.\\
The gravitational action for $f(T,\mathcal{T})$ gravity is given by
\begin{equation}
\label{7}
    S =\frac{1}{16\pi G} \int d^4x\, e[T+f(T,\mathcal{T})]+\int d^4x\, e\, \mathcal{L}_m
\end{equation}
Varying the action with respect to the vierbeins yields the field
equations
\begin{multline}
\label{8}
(1+f_T)\left[e^{-1}\partial_\mu (e e_{a}^{\,\,\alpha} S_{\alpha}^{\,\, \lambda \mu} )-e_{a}^{\,\, \alpha} T^{\mu}_{\nu \alpha} S_{\mu}^{\,\, \nu \lambda} \right] +  e_{a}^{\,\, \lambda}\left(\frac{f+T}{4}\right)+\\
 \left(f_{TT} \partial_{\mu}T+ f_{T\mathcal{T}}\partial_{\mu}\mathcal{T} \right) e_{a}^{\,\, \alpha} S_{\alpha}^{\,\, \lambda \mu} - f_{\mathcal{T}} \left(\frac{e_{a}^{\,\,\alpha} \overset{em}{T}_{\alpha}^{\,\, \lambda} + p_{m} e_{a}^{\,\, \lambda}}{2}\right)= \\
 4 \pi G e_{a}^{\,\,\alpha} \overset{em}{T}_{\alpha}^{\,\, \lambda}.
\end{multline}
where $f_\mathcal{T}={\partial f}/{\partial \mathcal{T}} $, $ f_{T \mathcal{T}}={\partial^2 f}/{\partial T\partial \mathcal{T}}$, and $\overset{em}{T}_{\alpha}^{\,\, \lambda}$ is the energy-momentum tensor.\\

We incorporate the flat FRW metric as usual to apply the aforementioned theory in a cosmological framework to  obtain modified Friedman equations. The FRW metric read
\begin{equation}
\label{9}
 ds^{2}=dt^{2}-a(t)^{2} \delta_{ij}dx^{i} dx^{j}, 
\end{equation}
where $a(t)$ is the scale factor. Further, \eqref{8} give rise to modified Friedmann equations:
\begin{equation}
\label{10}
H^2 =\frac{8\pi G}{3}\rho_m - \frac{1}{6}\left(f+12H^2f_T \right)+f_\mathcal{T}\left(\frac{\rho_m+p_m}{3} \right),
\end{equation}
\begin{multline} 
\label{11}
\dot{H}= -4\pi G(\rho_m+p_m)-\dot{H}(f_T-12H^2 f_{T \mathcal{T}})\\-H(\dot{\rho_m}-3\dot{p_m}) f_{T \mathcal{T }} - f_\mathcal{T}\left(\frac{\rho_m+p_m}{2} \right).
\end{multline}
Here, $\mathcal{T}=\rho_m-3p_m $ in the above equation is true for the perfect matter fluid.

Comparing the modified Friedmann equations \eqref{10} and \eqref{11} to General Relativity equations 
\begin{eqnarray}
\label{12}
H^2 &=& \frac{8 \pi G}{3}\left(\rho_m + \rho_{eff}\right),\\
\label{13}
\dot{H} &=& - 4\pi G \left(\rho_m + p_m + \rho_{eff}+p_{eff}\right).
\end{eqnarray}
we obtain 
                                                                                                                                                                                                                                                                                                          \begin{equation}
                                                                                                                                                                                                                                                                                                          \label{14}
                                                                                                                                                                                                                                                                                                          \rho_{eff} =\frac{1}{16\pi G}[f+12f_T H^2-2f_\mathcal{T}(\rho_m +p_m)]
                                                                                                                                                                                                                                                                                                          \end{equation}
                                                                                                                                                                                                                                                                                                          
                                                                                                                                                                                                                                                                                                          \begin{multline}
                                                                                                                                                                                                                                                                                                          \label{15}
                                                                                                                                                                                                                                                                                                          p_{eff} = \frac{1}{16\pi G}[f+12f_T H^2-2f_\mathcal{T}(\rho_m +p_m)]+ \\ (\rho_m+p_m)\left[\frac{(1+\frac{f_T}{8\pi G})} {1+f_T 12H^2 f_{TT}+H(\frac{d\rho_m}{dH})(1-3{c_{s}}^2)f_{T \mathcal{T}}} -1\right]
                                                                                                                                                                                                                                                                                                          \end{multline}
                                                                                                                                                                                                                                                                                                          
                                                                                                                                                                                                                                                                                                        The effective and total equation-of-state parameter is defined as follows;
                                                                                                                                                                                                                                                                                                        \begin{eqnarray}
                                                                                                                                                                                                                                                                                                        \label{16}
                                                                                                                                                                                                                                                                                                        \omega_{eff} &=& \frac{p_{eff}}{\rho_{eff}},\\
                                                                                                                                                                                                                                                                                                        \label{17}
                                                                                                                                                                                                                                                                                                        \omega &= & \frac{p_{eff} +p_m}{\rho_{eff}+\rho_m}.
                                                                                                                                                                                                                                                                                                        \end{eqnarray}
                                                                                                                                                                                                                                                                                                        We consider $p_m =0$ for the dust universe which implies $ \omega =\frac{\omega_{eff}}{1+\frac{\rho_m}{\rho_{eff}}}$.
                                                                                                                                                                                                                                                                                                        The conservation equation involving the effective energy and pressure reads   
                                                                                                                                                                                                                                                                                                         \begin{equation}
                                                                                                                                                                                                                                                                                                         \label{18}
                                                                                                                                                                                                                                                                                                         \dot{\rho}_{eff}+\dot{\rho_m}+3H(\rho_m+\rho_{eff}+p_m+p_{eff})=0.
                                                                                                                                                                                                                                                                                                         \end{equation}
                                                                                                                                                                                                                                                                                                         
                                                                                                                                                                                                                                                                                                        %   Finally, as an indicator of the accelerating dynamics of the Universe we use the deceleration parameter $q$, defined as;
                                                                                                                                                                                                                                                                                                          
                                                                                                                                                                                                                                                                                                        %   \begin{equation}\label{13}
                                                                                                                                                                                                                                                                                                        %   q=-\frac{\dot{H}}{H^2} -1
                                                                                                                                                                                                                                                                                                        %   \end{equation}
  
\section{Cosmology} \label{sec3}
 This section examines the cosmological impacts of $f(T,\mathcal{T})$ gravity while emphasizing on a specific model. We consider the functional form $f(T,\mathcal{T)}=\alpha\mathcal{T}+\beta T^2=\alpha \rho_m +\beta T^2 =\alpha \rho_m+ \gamma H^4 $, where  $\alpha $ and $\gamma=36\beta$  are constants \cite{Harko/2014}. For simplicity, we use $8 \pi G= c=1$. The model defines a straightforward deviation from GR inside the framework of $f(T,\mathcal{T)}$. The case $\alpha=0$, the model behaves as a  power-law cosmology in $f(T)$ theory \cite{Capozziello/2011}. In this case, we obtain $f_T=\frac{\gamma T}{18}$, $f_{TT} =\frac{\gamma}{18}$ ,$f_\mathcal{T}=\alpha$,  $f_{T \mathcal{T}}=0$.\\
 Hence, using the above expressions and equations \eqref{10} \& \eqref{11}, we have the following:
\begin{eqnarray}
\label{19}
\rho_{m} &=& \frac{3\left(1-\frac{\gamma H^2}{2} \right)}{1+\frac{\alpha}{2}} H^2, \\ 
\label{20}
\dot{H} &=& -\frac{3(1+\alpha) \left(1-\frac{\gamma H^2}{2}\right)}{(\alpha +2) (1-\gamma H^2)} H^2,\\
\label{21}
 q &=&\frac{3(1+\alpha)\left(1-\frac{\gamma H^2}{2}\right)}{(\alpha+2)(1-\gamma H^2)}-1.
\end{eqnarray}

Moreover, the effective dark energy density and pressure from equations \eqref{14} and \eqref{15} can be obtained as
\begin{eqnarray}
\label{22}
    \rho_{eff} &=&\frac{3H^2(\alpha+\gamma H^2)}{\alpha+2},\\
    \label{23}
    p_{eff}&=&-\frac{3H^2(\alpha+\gamma H^2)}{(\alpha+2)(\gamma H^2-1)}.
    \end{eqnarray}
which gives $ \omega_{eff}=\frac{1}{1-\gamma H^2}$.\\
Now, we replace the term $d/dt$ by $d/dlna$ via the expression $d/dt = H \frac{d}{dlna}$, ($a=\frac{1}{1+z}$) such that solution of equation \eqref{20} is
\begin{equation}
\label{24}
    H(z)= H_0 \sqrt{\frac{\sqrt{1-2^{-2 A} (2 z+2)^{2 A} \left(2 \gamma  H_{0}^2-\gamma ^2 H_{0}^4\right)}+1}{\gamma H_{0}^2}}
\end{equation}
where $A=\frac{3 (\alpha +1)}{\alpha +2}$.

\section{Observational constraints and Methodology} \label{sec4}
In this section, we will conduct a statistical study utilising the Monte Carlo Markov Chain (MCMC) approach, where we compare the predictions with data sets to the cosmic observations, to assess the viability of a model. In particular, we use Type Ia Supernovae (SNeIa) data, Baryon acoustic oscillation (BAO) data and the Observational Hubble data (H(z)).

\subsection{SNeIa data}
Since Type Ia Supernovae act as ``standard candles" and allow us to estimate cosmic distance. They are widely applied to impose constraints on the dark energy sector. In particular, we use the Pantheon compilation of 1048 points spanning the redshift range $0.01<z<2.26$ \cite{Scolnic/2018}. The $\chi^{2}$ function is given as 
\begin{equation}
\chi^2 _{SN}= \Delta\mu C^{-1}_{SN} \Delta\mu^{T},
\end{equation}
where $\Delta \mu= \mu_{i}-\mu_{th}$ is the difference between the observational and theoretical distance modulus, and $C^{-1}_{SN}$ corresponds to the inverse covariance matrix of the data. Further, we define $\mu= m_{B}-M_{B}$, where $m_{B}$ is the observed apparent
magnitude at a given redshift, while $M_{B}$ is the absolute
magnitude (Retrieving the nuisance parameters according to the new approach called BEAMS with Bias Correction (BBC) \cite{Kessler/2017}). The theoretical value is computed as 
\begin{eqnarray}
\mu_{th}&=& 5 log_{10}\left[\frac{d_{L}}{1 Mpc}\right]+25,\\
d_{L}&=& c(1+z) \int_{0}^{z} \frac{dy}{H(y,\theta)}.
\end{eqnarray}
where $\theta$ is the parameter space.

\subsection{Hubble data}
We make use of Hubble parameter measurements derived from the differential age method (often known as cosmic chronometer (CC) data). Here, we consider 31 points compiled in \cite{Moresco/2015}.
The $\chi^{2}$ function is given as 
\begin{equation}
\chi^2_{Hz}= \sum_{i=1}^{31} \frac{\left[H(z_{i})-H_{obs}(z_{i})\right]^2}{\sigma(z_{i})^{2}},
\end{equation}
where $H_{obs}$ is the observed value, $\sigma(z_{i})$ is the observational error.

\subsection{BAO}
Baryon acoustic oscillations (BAO) are pressure waves generated by cosmological perturbation in the baryon-photon plasma at the recombination epoch and appear as distinct peaks on large angular scales. We use BAO measurements from the Six Degree Field Galaxy Survey (6dFGS), Sloan Digital Sky Survey (SDSS), the LOWZ samples of Baryon Oscillation Spectroscopic Survey (BOSS) \cite{Blake/2011,Percival/2010}. The expressions used for BAO data are
\begin{eqnarray}
d_{A}(z) &=& c \int_{0}^{z} \frac{dz'}{H(z')},\\
D_{v}(z) &=& \left[\frac{d_{A}(z)^2 c z }{H(z)}\right]^{1/3},\\
\chi_{BAO}^2 &=& X^{T} C^{-1} X. 
\end{eqnarray}
Here, $d_{A}(z)$ is the comoving angular diameter distance, and $D_{v}(z)$ is the dilation scale and $C$ is the covariance matrix \cite{Giostri/2012}.

\subsection{Results}
The statistical results for the model is shown as contour plots in 
Figs. \ref{figure 1} and \ref{figure 2}. Also, table \ref{table1} corresponds to the values obtained for the parameter space to the combination of data sets. We noticed weaker constraints in case of $BAO$ whereas stronger constraints for $SNeIa$ and $joint (Hz+BAO+SNeIa)$ analysis. We assumed $\gamma=0.0006$ \cite{Harko/2014} so that $\frac{\gamma H^2}{2}<1$. We can observe that the BAO data is anti-correlated with other data sets. We also get $H_{0}$ constraint consistent with Planck results \cite{Planck/2018} favored by $\Lambda$CDM. 

\begin{widetext}

\begin{figure}[]
\centering
\includegraphics[scale=0.9]{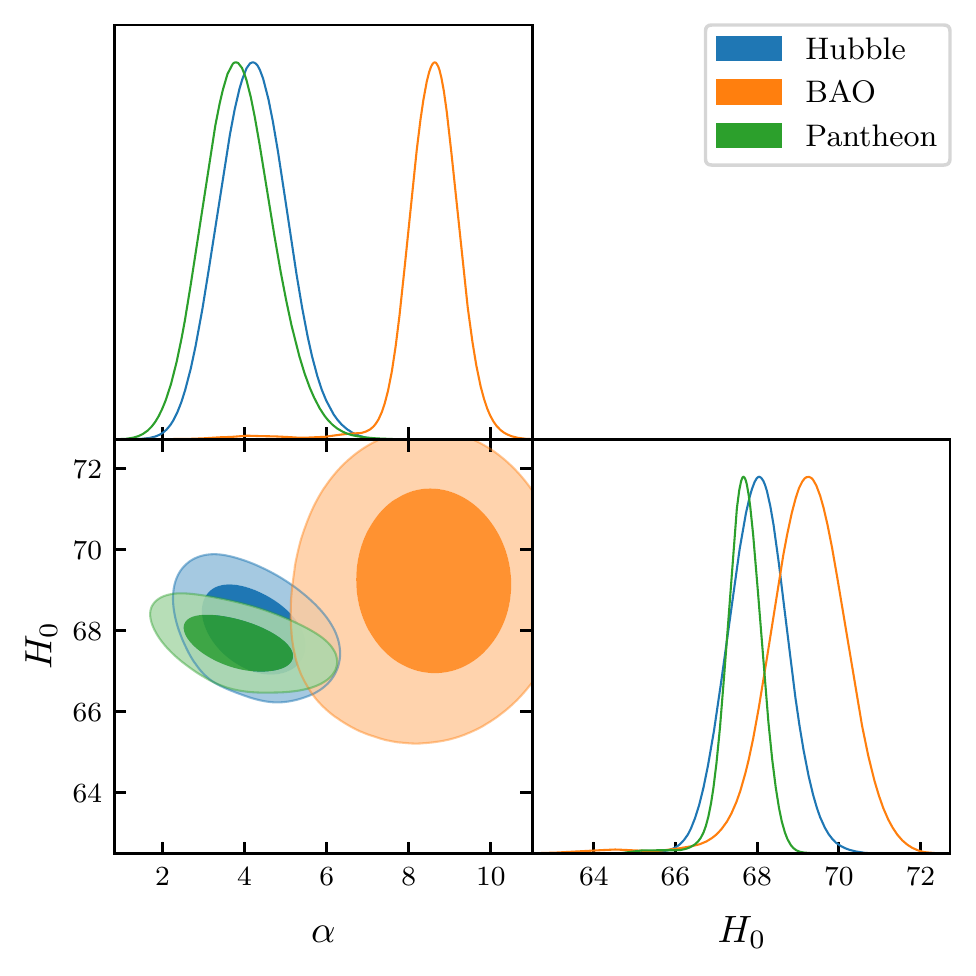}
\caption{One-dimensional and two-dimensional marginalized confidence regions (68\% CL and 95\% CL) for $\alpha$, $H_0$ obtained from the $Hubble$, $BAO$ and $Pantheon$ data for the f(T,$\mathcal{T}$) gravity model.}
\label{figure 1}
\end{figure}

\begin{figure}[]
\centering
\includegraphics[scale=0.9]{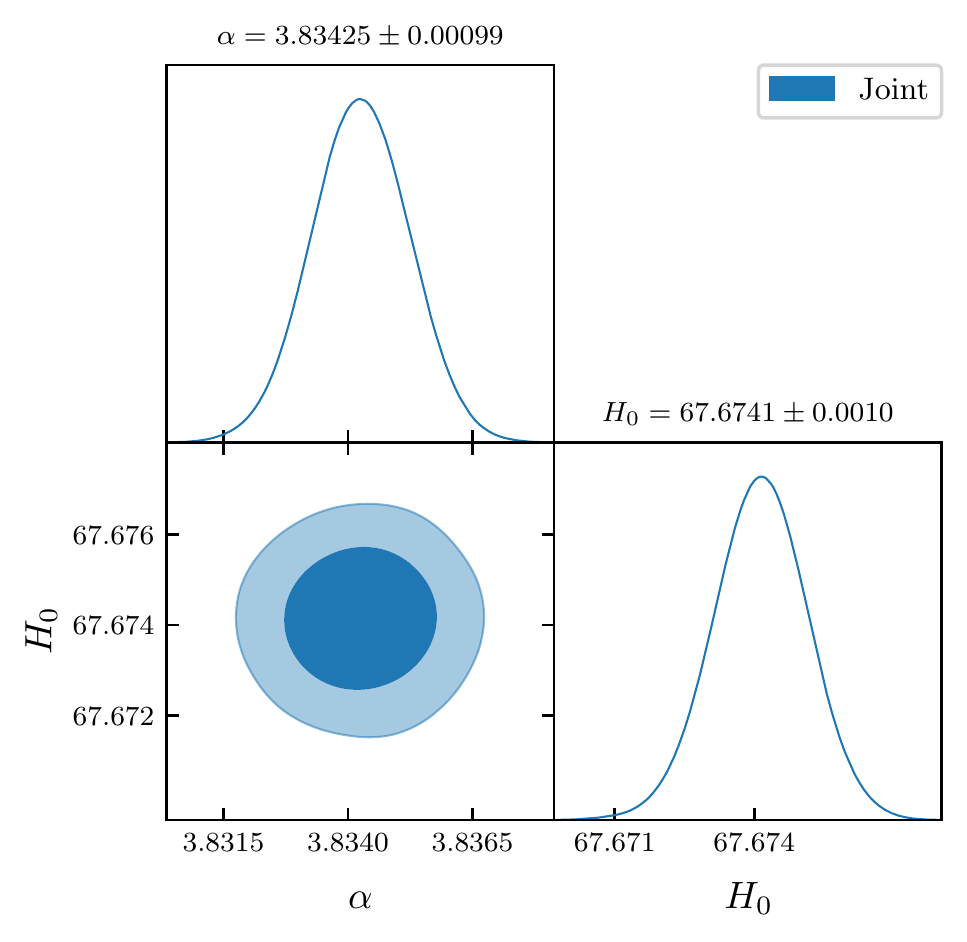}
\caption{One-dimensional and two-dimensional marginalized confidence regions (68\% CL and 95\% CL) for $\alpha$, $H_0$ obtained from the $Hubble+BAO+Pantheon$ data for the f(T,$\mathcal{T}$) gravity model.}
\label{figure 2}
\end{figure}

\begin{figure}[H]
\centering
\includegraphics[scale=0.5]{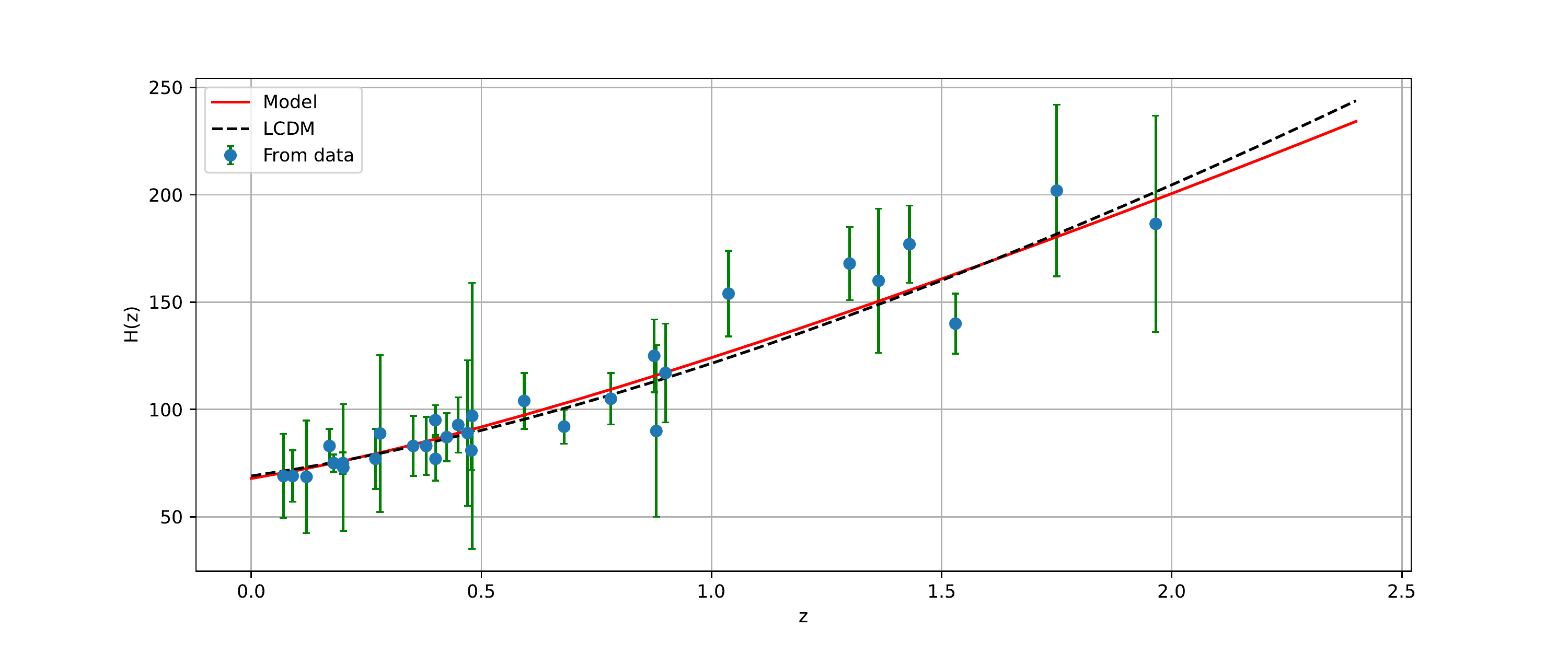}
\caption{Plot shows the expansion rate of the universe with theoretical predictions (red curve) and $\Lambda$CDM (black curve with $\Omega_{\Lambda_0}=0.7$ and $\Omega_{m_{0}}=0.3$). The blue dots represent 31 Hubble points with the corresponding error bars.}
\label{figure error}
\end{figure}

\begin{figure}[H]
\centering
\includegraphics[scale=0.5]{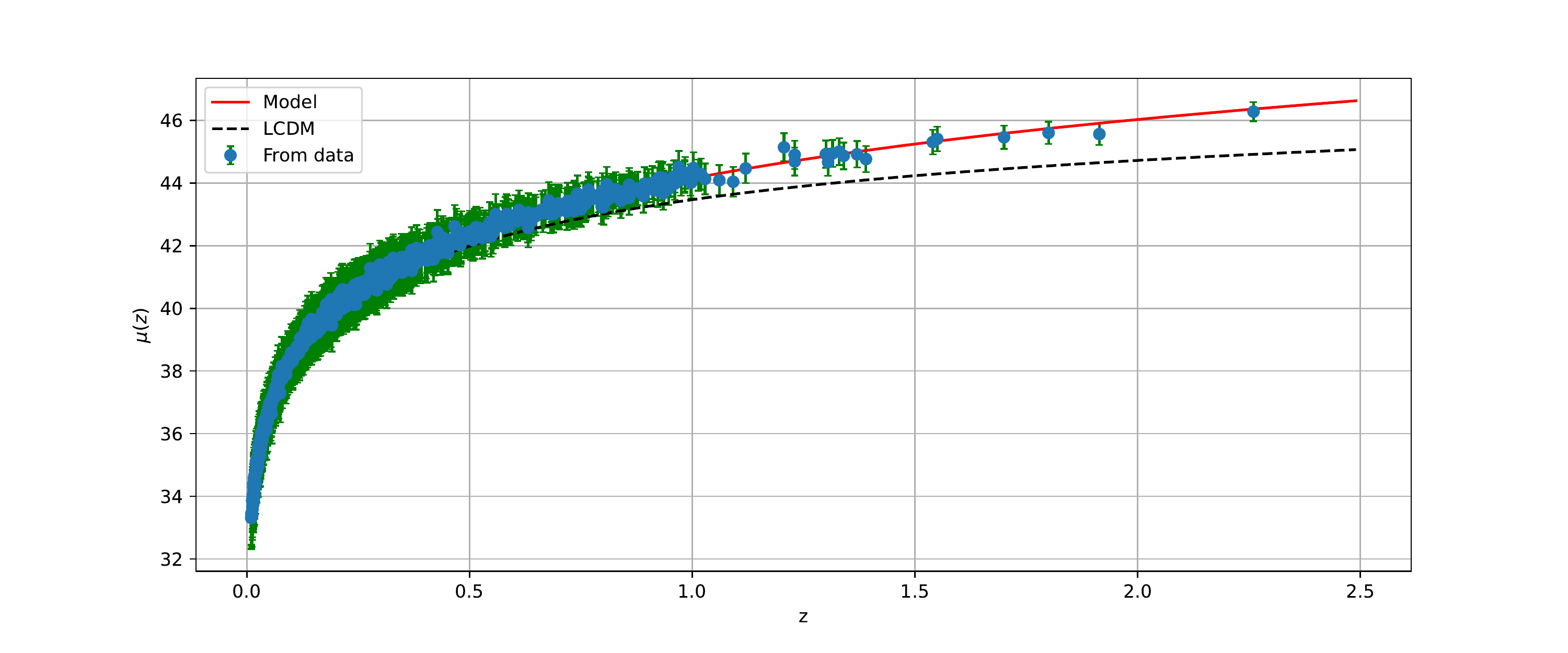}
\caption{Plot shows the $\mu$-redshift relation of Pantheon SN sample  with theoretical predictions (red curve) and $\Lambda$CDM (black curve). with $\Omega_{\Lambda_0}=0.7$ and $\Omega_{m_{0}}=0.3$). The blue dots represent 1048 Pantheon points with the corresponding error bars.}
\label{figure mu}
\end{figure}

\begin{table}[H]
\begin{center}
 \caption{Best-fit values of model parameters obtained from observational datasets}
    \label{table1}
\begin{tabular}{|l|c|c|c|c|c|}
\hline 
Datasets              & $\alpha$ & $H_{0}$ & $z_{t}$ & $q_{0}$ & $\omega_{0}$\\
\hline

$BAO$           & $8.55^{+0.63}_{-0.53}$  & $69.20^{+1.0}_{-1.0}$ &  $0.36^{+0.04}_{-0.04}$ & $-0.36^{+0.03}_{-0.04}$ & $-0.57^{+0.024}_{-0.024}$\\
\hline
$Hz$     & $4.21^{+0.82}_{-0.82}$  & $68.01^{+0.71}_{-0.71}$  & $0.60^{+0.141}_{-0.101}$ & $-0.45^{+0.04}_{-0.04}$ &  $-0.63^{+0.024}_{-0.026}$ \\
\hline
$SNeIa$     & $3.85^{+0.79}_{-0.91}$  & $67.66^{+0.48}_{-0.48}$  & $0.65^{+0.198}_{-0.092}$ &  $-0.46^{+0.02}_{-0.04}$& $-0.64^{+0.019}_{-0.024}$\\
\hline
$Hz+BAO+SNeIa$             & $3.83^{+0.0009}_{-0.0009}$  & $67.67^{+0.0010}_{-0.0010}$ & $0.65^{+0.0004}_{-0.0017}$ &  $-0.46^{+0.0001}_{-0.937}$ & $-0.64^{+0.00003}_{-0.00003}$\\ 
\hline
\end{tabular}
\end{center}
\end{table}

\end{widetext}

\section{Cosmological Evolution} \label{sec5}
\begin{figure}[]
\centering
\includegraphics[scale=0.5]{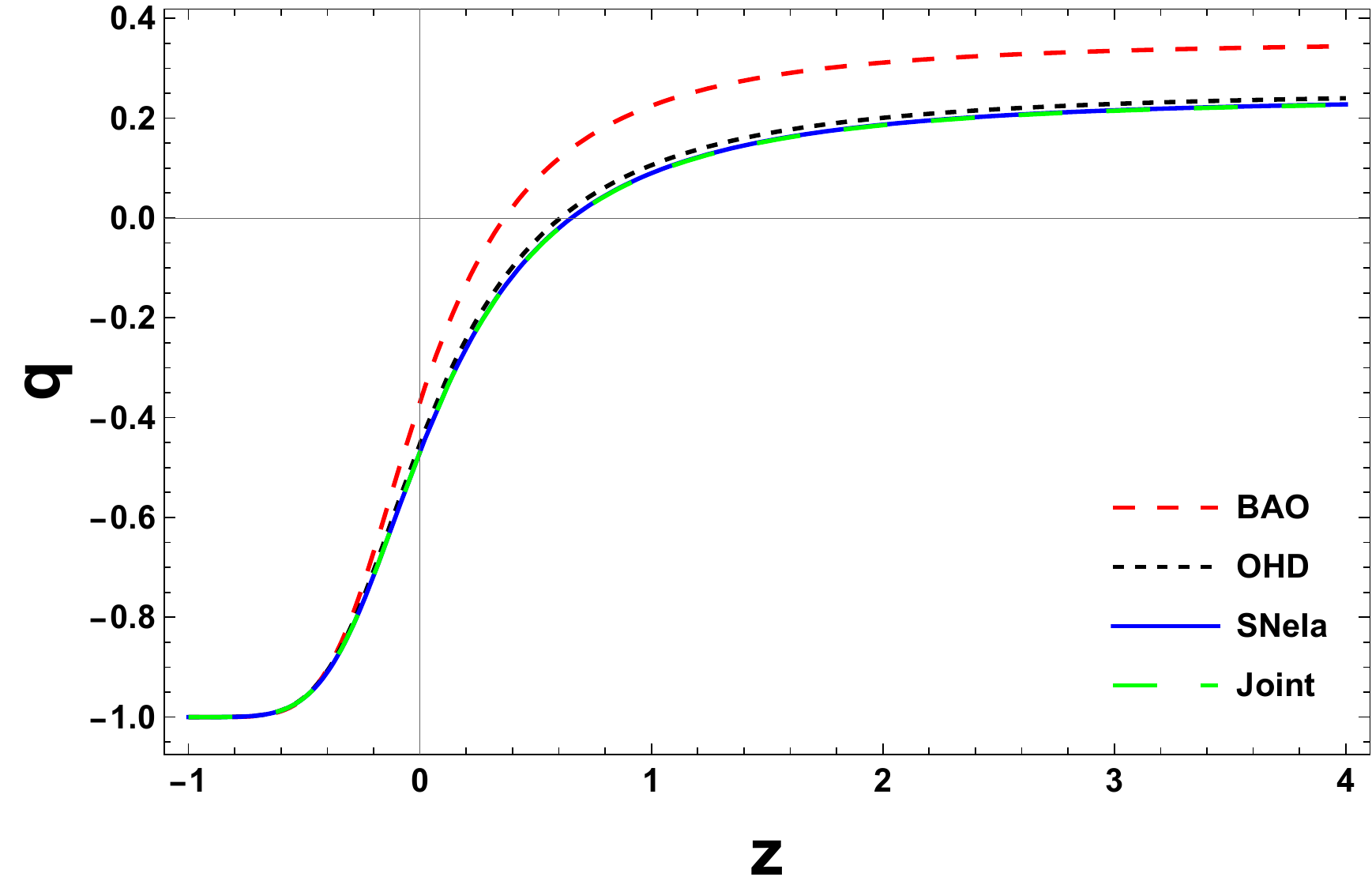}
\caption{Variation of the deceleration parameter $q$ as a function of the redshift $z$ for different data sets.}
\label{figure 4}
\end{figure}

The plots in this section demonstrate how the universe can have very intriguing dynamics depending on the values of the parameters. 
The Hubble function, presented in figure \ref{figure error} is a monotonically increasing function of redshift throughout the entire evolution of the universe.\\
Figure \ref{figure 4} shows that the universe begins its history from deceleration ($q>0$) and shows accelerating phase ($q<0$) after a transition redshift $z_{t}$. The deceleration parameter is defined as $q= -\frac{\dot{H}}{H^2}-1$. This evolution is consistent with the recent universe behaviour as it went through three stages: a decelerating dominated phase, an accelerating expansion phase, and a late-time accelerating phase. Keep in mind that the universe terminates in a de Sitter expansion at asymptotically lower redshifts. We find that the present value of deceleration parameter ($q_{0}$) \cite{Almada/2019,Basilakos/2012}  and $z_{t}$ \cite{Garza/2019,Jesus/2020} is in good agreement with $SNeIa$ and $Hz+BAO+SNeIa$ data sets.

\begin{figure}[]
\centering
\includegraphics[scale=0.5]{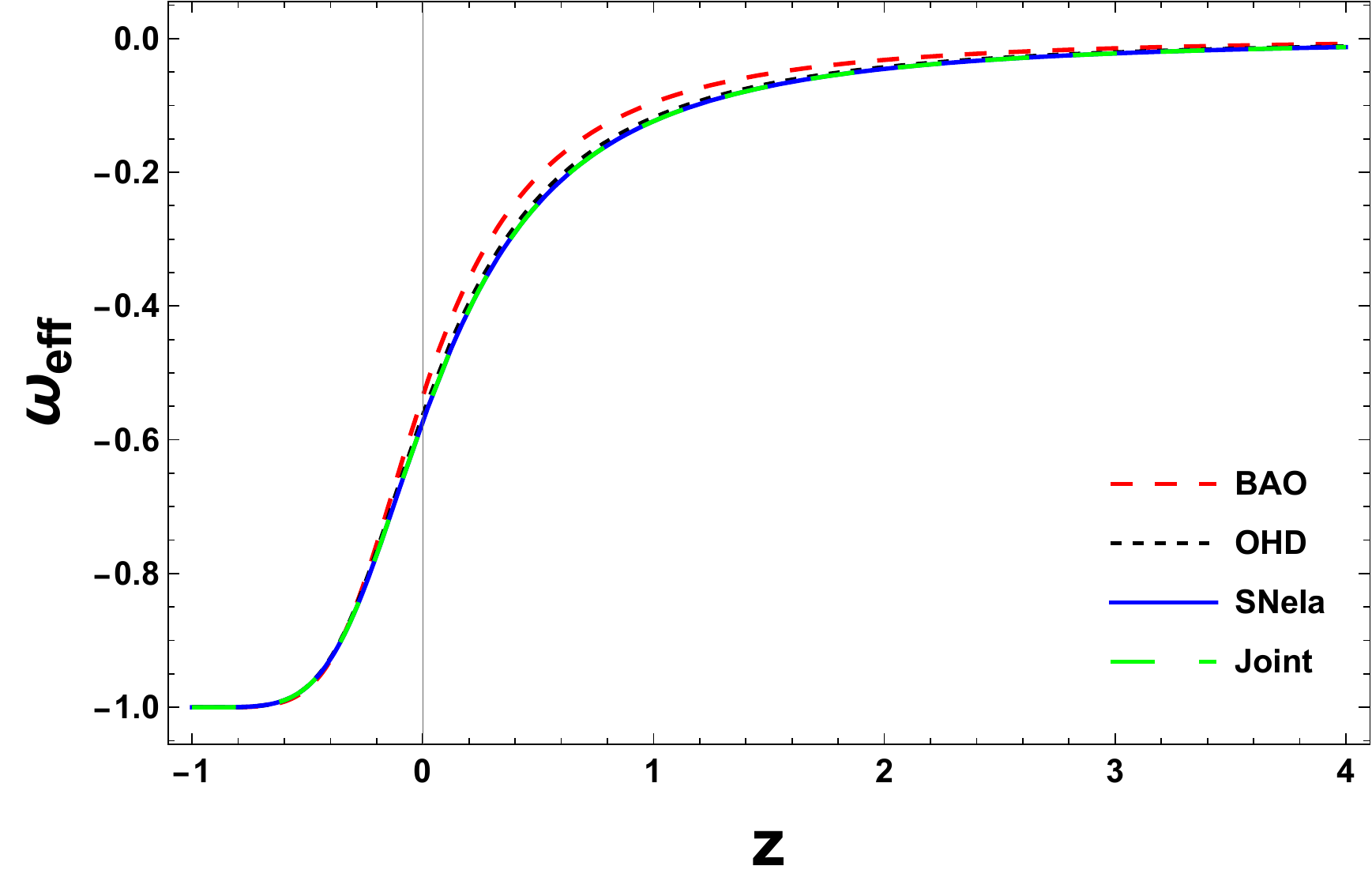}
\caption{Variation of $\omega_{eff}$ as a function of the redshift $z$ for different data sets}
\label{figure 5}
\end{figure}

\begin{figure}[]
\centering
\includegraphics[scale=0.5]{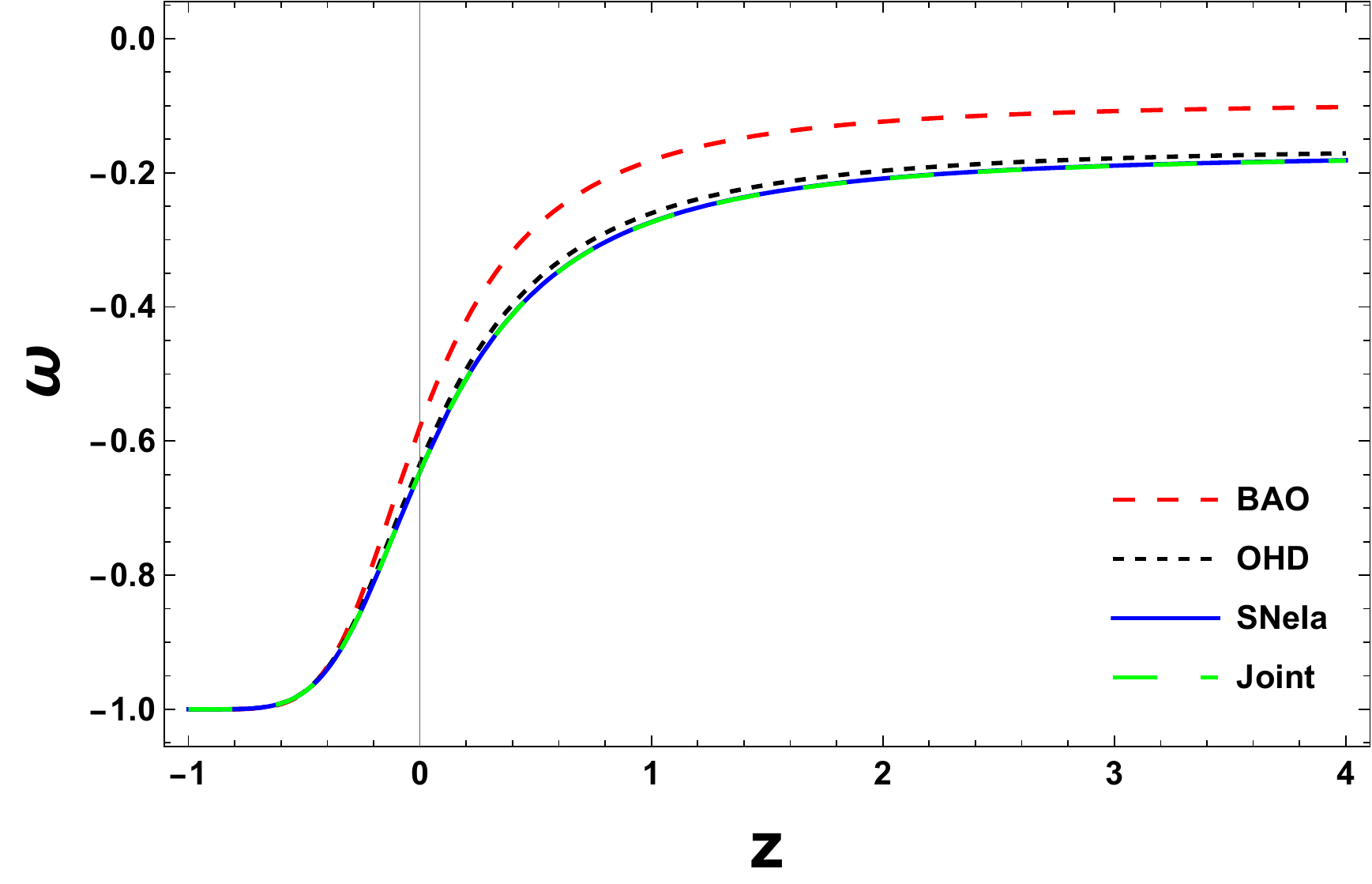}
\caption{Variation of $\omega$  as a function of the redshift $z$ for different data sets}
\label{figure 6}
\end{figure}

Determining the equation of state's value and its evolution is another attempt to comprehend the existence of dark energy. 
The equation of state ($\omega_{eff}$) in figure \ref{figure 5} show a similar evolution,  moving towards negative at lower redshifts. Moreover, we show  the total equation of state parameter ($\omega$) in figure \ref{figure 6}. Hence, the  both the equation of state parameter lie in the quintessence regime ($-1<\omega<0$), approaching the cosmological constant ($\omega=-1$) at smaller redshifts. We find that the present value of $\omega_{0}$ is in good agreement with $SNeIa$ and $Hz+BAO+SNeIa$ data sets.

\section{Conclusion} \label{sec6}
Inspired by the teleparallel-formulation of general relativity, we attempted to investigate the extension of $f(T)$ gravity based on the coupling between the torsion scalar $T$ and the trace of energy-momentum tensor $\mathcal{T}$. The essential point is that both $f(T)$ components, as well as the matter energy density and pressure, contribute to the effective dark energy sector. The additional freedom of the imposed Lagrangian in the $f(T,\mathcal{T})$ cosmology allows for a very wide range of conditions and  behaviors.\\

In the current study, we investigated the cosmological implications of $f(T,\mathcal{T})$ theory. We considered the squared-torsion model $f(T,\mathcal{T})= \alpha \mathcal{T}+ \beta T^2$, where $\alpha$ and $\beta$ are free parameters. We obtained the solution of modified Friedmann equations in the form of Hubble parameter as a function of redshift $z$. Further, in section \ref{sec3}, we  employed the recent observational data: $Hubble$, $BAO$, $SNeIa$ and the $joint$ analysis to constrain the unknown model parameters. In figures \ref{figure 1} and \ref{figure 2}, we obtained the best-fit values of model parameters. In comparison with the $\Lambda$CDM model, the obtained $H(z)$ and the $\mu(z)$ of the considered model are confronted to the cosmic data in figures \ref{figure error} and \ref{figure mu} respectively. \\

Depending on the model parameters constrained, we discovered a wide range of intriguing cosmological behaviors. For instance, we found evolution of deceleration parameter, explicitly experiencing a change from a deceleration to acceleration, capable of explaining the late-time universe. Additionally, the effective EoS ($\omega_{eff}$) and total EoS ($\omega$) behaves in a similar fashion demonstrating that the cosmic fluid has the characteristics of quintessence dark energy. Moreover, we find that the
present values of $q_{0}$, $\omega_{0}$ and $z_{t}$ are in good agreement with $SNeIa$ and $Hz + BAO + SNeIa$ data sets.\\
Finally, it is essential to point that $f(T,\mathcal{T})$ subject to observational data can explain late-time accelerating universe and can be applied to different regimes to establish a viable gravitational formalism. Furthermore, the perturbation analysis could  be extended to the vector and tensor analysis which is useful in predicting the inflationary scenario. We hope that this analysis will encourage readers to consider torsional modified gravity as a candidate to describe the universe.

\acknowledgments  
SA acknowledges BITS-Pilani, Hyderabad Campus for the financial support. AB acknowledges University Grants Commission (UGC) Maulana Azad National Fellowship (MANF), New Delhi, India for awarding Junior Research Fellowship (UGC-Ref.No.: 211610222082). SA \& PKS acknowledge IUCAA, Pune, India for providing support through the visiting Associateship program.

%\newpage
%\section{References}

\end{document}